\documentclass{JHEP3}
\usepackage{epsfig,psfig}
\usepackage{amsfonts,amssymb,latexsym}
\newcommand{\beq}{\begin{eqnarray*}}
\newcommand{\eeq}{\end{eqnarray*}}
\newcommand{\beqn}{\begin{eqnarray}}
\newcommand{\eeqn}{\end{eqnarray}}

\newcommand{\Z}{\mathbb{Z}}

\newcommand{\R}{\mathbb{R}}
\newcommand{\C}{\mathbb{C}}

\newcounter{fig}

\newcounter{gif}

\newcommand{\mref}[1]{$(\ref{#1})$}
\newcommand{\pa}{\partial}

\def\sqr#1#2{{\vcenter{\vbox{\hrule height.#2pt
            \hbox{\vrule width.#2pt height#1pt \kern#1pt
                  \vrule width.#2pt}\hrule height.#2pt}}}}

\newcommand{\Tc}{{\cal T}}

\newcommand{\zb}{\bar{z}}
\newcommand{\xb}{\bar{x}}
\newcommand{\yb}{\bar{y}}
\newcommand{\wb}{\bar{w}}
\newcommand{\vb}{\bar{v}}
\newcommand{\hb}{\tilde{h}}
\newcommand{\xib}{\bar{\xi}}
\newcommand{\etab}{\bar{\eta}}
\newcommand{\bxi}{\bar{\xi}}
\newcommand{\zetab}{\bar{\zeta}}

\newcommand{\corr}[1]{\langle \, #1 \, \rangle}
\newcommand{\bcorr}[1]{\bigg\langle \, #1 \, \bigg\rangle}
\newcommand{\cet}[1]{\mid \! #1 \, \rangle}
\newcommand{\bet}[1]{\mid \! #1 \! \mid}

\renewcommand{\d}{\mbox{d}}

\newcommand{\Gt}{\tilde{G}}

\newcommand{\eps}{\epsilon}

\title{Twistfield Perturbations of Vertex Operators in the $\Z_2$-Orbifold Model}
\author{Holger Eberle\\ 
Physikalisches Institut der Universit\"at Bonn, Nu\ss allee 12, 53115 Bonn, Germany\\
E-mail: \email{eberle@th.physik.uni-bonn.de}}

\abstract{We apply Kadanoff's theory of marginal deformations of conformal field theories
to twistfield deformations of $\Z_2$ orbifold models in K3 moduli space. 
These deformations lead away from the $\Z_2$ orbifold sub-moduli-space
and hence help to explore conformal field theories which have not yet been understood.
In particular, we calculate
the deformation of the conformal dimensions of vertex operators for $h,\hb <1/2$ in
second order perturbation theory.}


\preprint{BONN-TH-2001-02\\ hep-th/0103059}

\begin{document}








\section{Introduction}

Orbifolds of torus models are a very promising starting point for the study of
more complicated conformal field theories. The reason for that is that their structure is
still very close to the one of their ``mother theory'', the torus model, which
is known very well.
The whole untwisted sector of the theory is inherited from the torus model, only
the behaviour of the twistfields provides quite some difficulties
to understand.
However, especially in the case of the abelian
$\Z_N$ orbifolds, a lot of progress has already been made in understanding 
the conformal properties of and with twistfields, e.g.\ in \cite{BR87}, \cite{BR87b},
\cite{HV87}, \cite{DFMS87}, \cite{Dix87}, \cite{Zam87}, \cite{ADG88}, 
\cite{BKM91}, \cite{EJLM92}, \cite{Jun92}.

Now, our immediate interest lies in the K3 part of the moduli space of N=(4,4)
superconformal theories with central charge c=(6,6), although the following considerations
and calculations might as well be generalised to other moduli spaces with
N=(2,2) supersymmetry. This part of the moduli space contains quite a few Gepner models
and submoduli spaces corresponding to cyclic orbifolds of the N=(4,4) supertoroidal
theories with central charge c=(6,6), but neither does it contain toroidal theories themselves
nor is there anything known about the theories in the vast empty spaces between
these known subvarieties (see e.g. \cite{NW99}, \cite{Wend00}).

The idea is to use the theory of deformations of conformal field theories developed by
Kadanoff already in 1978 (\cite{Kad78}, \cite{BK79}) in order to gain
information about theories lying close to known subvarieties in moduli space.
Kadanoff's theory gives a prescription how to calculate a correlator in a theory which
is deformed away from a known theory by an exactly marginal operator in terms of integrals over
correlation functions in the known theory. 
As a conformal field theory in this moduli space is completely determined 
by the conformal weights and the 
three point correlation functions of its fields, it is sufficient to study their
behaviour under deformations.

The aim of this article is to study the deformation of the conformal weights of
vertex operators with small conformal weights $h, \hb <1/2$ in directions corresponding
to marginal twistfields in second order
perturbation theory. We will carry out this calculation starting off from any point
of the 16-dimensional $\Z_2$ orbifold hyperplane in the above mentioned K3 part of moduli space.

The outline of the article is as follows. First we derive the
four point correlator, which we need for our calculation. 
This part contains quite a long and tedious calculation the result of which is given in
\mref{eigentlich2}.
It follows
a discussion of the singularities of this correlator and the necessary
renormalisation procedure. Then, we present the numerical results of
the integration, analyse these and show the implications for the
moduli space of K3.


\section{The correlator\label{vertex1}}

In the following, we want to investigate certain deformations of $\Z_2$ orbifold theories at central charge
$c=6$. These are unitary, superconformal orbifolds of toroidal theories with orbifold group
$\Z_2$. The following correlators are all calculated in a certain $\Z_2$ orbifold model,
if not denoted otherwise.
According to Kadanoff \cite{Kad78}, a deformation 
of such a theory is given by the addition of exactly marginal operators $O_j (w,\wb)$
to the action of the original theory. A general deformation of the action is given by 
$\int d^2w \; \sum_{j\in I} g_j O_j (w,\wb)$, where $I$ is the index set of all exact marginal operators
and $g_j \in \R$ are some (suitably small) coupling coefficients.

Now, the two point function of two conformal fields is completely determined by the
conformal weights of these fields (up to normalisation). Hence, the deformation of conformal weights
is given by the deformation of the two point function of a field with itself. This leads
Kadanoff to the first order calculation \cite{Kad78}
\beqn
\label{direkt}
\lefteqn{\!\!\!\!\!\!\!\!\!\frac{\pa}{\pa g_i} \; \corr{ \Phi_{\alpha} (x,\xb) \; \Phi_{\alpha} (0,0)}_{\bf g} \; 
\Big|_{{\bf g} = 0} } \nonumber \\
&&= \left(  -2 \frac{\pa h(g_i)}{\pa g_i}\log (x) -2\frac{\pa \hb(g_i)}{\pa g_i} \log (\xb) \right)  \;
\corr{ \Phi_{\alpha} (x,\xb) \; \Phi_{\alpha} (0,0)}_0 \; \nonumber \\
&&= \int \d^2w \; \corr{O_i (w,\wb) \Phi_{\alpha} (x,\xb) \; \Phi_{\alpha} (0,0) }_0 \; ,
\eeqn
where the $O_i (w,\wb)$ denote the exactly marginal operators (as above),
$\corr{ . }_{\bf g}$ the correlator in the deformed theory (deformed by  $\sum_{j\in I} g_j O_j (w,\wb)$,
${\bf g}$ denotes the vector of the $g_j$)
and $h, \hb$ the conformal weights of 
$\Phi_{\alpha}$ w.r.t.\ the left resp.\ right Virasoro field.
But, taking $O_i (w,\wb)$
to be an exactly marginal twistfield and $\Phi_{\alpha} (x,\xb)$ to be a bosonic
vertex operator, this order of perturbation theory naturally vanishes due to the
point group selection rule. The point group selection rule requires that all twists of the
correlator sum up to an integer number. In the case of the orbifold group $\Z_2$ that means
that we always need an even number of twistfields in a correlator. (See e.g.\ \cite{Dix87} 
for an introduction to conformal field theory on cyclic orbifolds.) 
However, as long as all orders in perturbation
theory have vanished up to the $n^{\mbox{\tiny th}}$ order, the result in
\mref{direkt} can easily be generalised to the $(n\!+\!1)^{\,\mbox{\tiny st}}$ order
\beqn
\label{direkt2}
\lefteqn{\!\!\!\!\!\!\!\!\!\frac{\pa^{n+1}}{\pa g_i^{n+1}} \; \corr{ \Phi_{\alpha} (x,\xb) \; 
\Phi_{\alpha} (0,0)}_{\bf g} \; \Big|_{{\bf g} = 0} = } \nonumber \\
&=& \left(  -2 \frac{\pa^{n+1} h}{\pa g_i^{n+1}}\log (x) -2\frac{\pa^{n+1} \hb}{\pa g_i^{n+1}} 
\log (\xb) \right)  \; \corr{ \Phi_{\alpha} (x,\xb) \; \Phi_{\alpha} (0,0)}_0 \nonumber \\
&=&\int \d^2w_1 \ldots \int \d^2w_{n+1} \; \corr{ \Phi_{\alpha} (x,\xb) \; \Phi_{\alpha} (0,0)
O_i (w_1) \ldots O_i (w_{n+1})}_{0,c} \;\; ,
\eeqn
where the last correlator only contains the connected part, i.e.\ all contributions
to this correlator originating in a factorisation of the correlator into two or more correlators
have to be subtracted, and where the integral
has to be suitably regulated by a cut-off.
Hence, we can proceed to calculate the second order perturbation of
the conformal weights by calculating the corresponding 
four point correlation function first.

The four point correlation function of two vertex operators with two ground state twistfields
in $\Z_2$ orbifolds of toroidal theories
has been known for quite a while (see e.g. \cite{HV87}, \cite{EJLM92}, \cite{Jun92};
a short derivation can be found in the appendix).
However, doing perturbation theory with twistfields, the proper marginal operators to perturb
a correlation function are not the (anti-)chiral
ground state twistfields in the NSNS sector with conformal dimension 
$h, \hb =1/2$ but their superpartners.
Only these have the right conformal dimensions $h, \hb =1$ and are exactly marginal
to all orders in perturbation theory \cite{Dix87}.

We take the $\Z_2$ orbifold group to act on the holomorphic U(1) currents $j_s (z)= i\partial X_s$ of the 
underlying toroidal theory by multiplication with a minus sign, i.e.\ $j_s (z) \mapsto -j_s (z)$. 
Hence, the vector of U(1) charges $p$ of the torus vertex operators $V^t_p$ w.r.t.\ to these toroidal
U(1) currents is also multiplied with a minus sign under this group action. The vertex operators are
mapped $V^t_p \mapsto V^t_{-p}$. The superscript $t$ signifies a field of the original torus theory.
Thus, the $\Z_2$ orbifold model only contains the symmetrical linear combinations of torus vertex operators
\beq
V_{p} = \frac{1}{\sqrt{2}} \left( V^t_p + V^t_{-p} \right) \; ,
\eeq
as these are invariant under the group action. As we always work in even real dimensions, we will often use 
the linear combinations $j^{(k)}_{\pm}=\frac{1}{\sqrt{2}}(j_k \pm i j_{k+d/2})$ for
the bosonic currents of the torus theory, and for
the charges $p \equiv (p^{(k)}_+ , p^{(k)}_- )_{k=1\dots d/2}$ respectively. 
The upper index in brackets $k=1 \dots d/2$ signifies the complex dimension,
the lower index $s=1\dots d$ the real dimension. The normalisation
of the $V_p^t$ is fixed in the appendix.  

In order to build correct marginal operators
for a unitary deformation of the toroidal theory, we first take the
hermitian linear combinations of ground state twistfields 
$\frac{1}{\sqrt{2}} (T_{+1/2}^f + T_{-1/2}^f)$ where $f$ signifies the fixed point at which the respective
field is localised, where
the $\pm$ sign indicates whether this field belongs to an (anti)twist and where $k/N=1/2$ gives the
corresponding twist itself. (For an introduction to cyclic orbifolds and the fields in the twisted sectors
we refer to e.g.\ \cite{Dix87}.)
We find the proper exactly marginal twistfields
as their (left+right) superpartners with conformal dimensions $h, \hb =1$, represented by
\beq
S^f=\frac{1}{\sqrt{2}} (S_{+1/2}^f + S_{-1/2}^f) \; .
\eeq
For ease of notation, we will omit the superscript $f$ in the following; it is understood that all twistfields
live at the same fixed point.

Now, we can start to calculate the correlator
\beqn
\label{eigentlich}
Z &\equiv& 
\corr{V_{p} (z,\zb)\; V_{-p} (w,\wb)\; S (x,\xb)\; S (y,\yb)} \nonumber \\
&=& \corr{V_{p} (z,\zb)\; V_{-p} (w,\wb)\; S_{+1/2} (x,\xb)\; S_{-1/2} (y,\yb)} \nonumber \\
&=& \bcorr{V_{p} (z,\zb)\; V_{-p} (w,\wb)\; \left(\oint_x \d \xi \; G^-(\xi) \oint_{\xb} 
\d \xib \; \Gt^-(\xib) \; T_{+1/2} (x,\xb)\; \right) \nonumber \\
&& \left(\oint_y \d \eta \; G^+(\eta) \oint_{\yb} 
\d \etab \; \Gt^+(\etab) \; T_{-1/2} (y,\yb) \right) } \; 
\eeqn
as part of the desired correlator. The first equality follows because the correlator with 
$S_{+1/2}$ and $S_{-1/2}$ interchanged will turn out to give
exactly the same result as the one in the second line above and because 
these are the only two non-vanishing terms by
the point group selection rule. In the third line, we use the fact that the $S_{\pm 1/2}$ are the 
superpartners of the ground state twistfields $T_{\pm 1/2}$, and hence can be constructed by applying the
suitable supercurrents $G^{\pm}(\eta)$, $\Gt^{\pm}(\eta)$ of the N=(2,2)
Super--Virasoro--Algebra to the ground state twistfields. Notice that $T_{+ 1/2}$ is a chiral/chiral primary,
$T_{- 1/2}$ an antichiral/antichiral primary field.

There are two important things to notice. First, as we are dealing with a correlator
containing only two twistfields, there are no classical solutions to be taken into
account. Hence, the holomorphic and the antiholomorphic part of the correlator
factorise. We will therefore concentrate
on the holomorphic part in the following. Second, performing the 
supersymmetry transformations on the ground state twistfields themselves we run
into deep problems as excited twistfields turn up whose correlation functions are rather nasty
to handle and which have not been studied much so far. Hence, the idea is to take the contours 
of the integrals in \mref{eigentlich} not around a single twistfield but around all other fields in the
correlator instead.  This 'turning around of the contour' gives a negative sign. As the supercurrents
$G^{\pm}$ drop off as $1/z^3$ at infinity, there is no contribution at infinity. Now,
taking into account the negative sign when interchanging two fermionic fields
and noticing that there are only finite contributions in the OPE of $G^+(\eta)$ and 
the chiral primary field $T_{+1/2}(x)$, we get for the holomorphic factor
\beqn
\label{lang}
\tilde{Z} &\equiv& \bcorr{V_{p} (z)\; V_{-p} (w)\;  
\left(\oint_x \d \xi \; G^-(\xi)\; T_{+1/2} (x)\right) \; 
\left(\oint_y \d \eta \; G^+(\eta) \; T_{-1/2} (y)\right)}  \nonumber \\
&=& \frac{1}{2} \bigg[\bcorr {\left(\left(\oint_z \d \xi \; G^-(\xi)\;\oint_z \d \eta \; G^+(\eta) 
- \oint_z \d \eta \; G^+(\eta) \; \oint_z \d \xi \; G^-(\xi) \right) V_{p} (z)\right) 
\; \nonumber \\
&& V_{-p} (w)\; T_{+1/2} (x)\; T_{-1/2} (y)} \nonumber \\
&& + \bcorr {\left( V_{p} (z) \left(\oint_w \d \xi \; G^-(\xi)\;\oint_w \d \eta \; G^+(\eta) 
- \oint_w \d \eta \; G^+(\eta) \; \oint_w \d \xi \; G^-(\xi) \right) V_{-p} (w)\;\right) \nonumber \\
&& \; T_{+1/2} (x)\; T_{-1/2} (y)} \nonumber \\
&& + \bcorr{V_{p} (z)\; V_{-p} (w)\; \left( \oint_x \d \eta \; G^+(\eta) \; \oint_x \d \xi \; G^-(\xi)
T_{+1/2} (x) \right)\; T_{-1/2} (y)} \nonumber \\
&& - \bcorr{V_{p} (z)\; V_{-p} (w)\; T_{+1/2} (x) 
\left( \oint_y \d \xi \; G^-(\xi) \; \oint_y \d \eta \; G^+(\eta) \; T_{-1/2} (y)\right)} \nonumber \\
&& + 2 \; \bcorr {\left(\oint_z \d \xi \; G^-(\xi) V_{p} (z)\right) 
\left( \oint_w \d \eta \; G^+(\eta) \; V_{-p} (w)\;\right)  \; T_{+1/2} (x)\; T_{-1/2} (y)} \nonumber \\
&& - 2 \;  \bcorr {\left(\oint_z \d \eta \; G^+(\eta) V_{p} (z)\right) 
\left( \oint_w \d \xi \; G^-(\xi) \; V_{-p} (w)\;\right)  \; T_{+1/2} (x)\; T_{-1/2} (y)} \bigg]  .
\eeqn

First, we look at the first two terms of \mref{lang}. We define $\psi_{\pm}^{(i)}$ to be the fermionic
superpartners of the bosonic currents $j_{\pm}^{(i)}$, introduced earlier, where $i=1\dots d/2$ with 
$d$ the real dimension, hence $d/2$ the complex dimension. The normalisation is given by the OPEs
\beq
\psi^{(i)}_+ (z) \; \psi^{(j)}_-(w) \sim \frac{\delta^{ij}}{z-w} \quad ,  \quad \quad
j^{(i)}_+ (z) \; j^{(j)}_-(w) \sim \frac{\delta^{ij}}{(z-w)^2} \; .
\eeq
Then, using the explicit expression for the supercurrents (see e.g. \cite{Gre96})
\beqn
\label{super}
G^{\pm} (z) &=& \sqrt{2} \; \sum_{i=1}^{d/2} :\psi_{\pm}^{(i)} j_{\mp}^{(i)}: (z) \; ,
\eeqn
we can calculate ($p^2=\sum_{k=1}^{d} p_k p_k = \sum_{i=1}^{d/2} (p_+^{(i) \; 2} + p_-^{(i) \; 2})$) 
\beqn
\label{eins}
\lefteqn{\bcorr {\left(\oint_z \d \xi \; G^-(\xi)\;\oint_z \d \eta \; G^+(\eta)  V_{p} (z)\right) 
\; V_{-p} (w)\; T_{+1/2} (x)\; T_{-1/2} (y)} =} \nonumber \\
&=& \bcorr {\left(\oint_z \d \xi \; G^-(\xi)\;\sqrt{2} \sum_{i=1}^2 \;p_-^{(i)} \; :\ \psi_+^{(i)} 
 \frac{1}{\sqrt{2}} \left( V^t_p - V^t_{-p} \right): (z)\right) 
\; \nonumber \\
&&  V_{-p} (w)\;T_{+1/2} (x)\; T_{-1/2} (y)} \nonumber \\
&=& \bcorr {\bigg(2 \sum_{j=1}^2 p_-^{(j)}\; \frac{1}{\sqrt{2}}: (j_{j} + ij_{j+d/2}) 
\frac{1}{\sqrt{2}} \left( V^t_p - V^t_{-p} \right): (z) \nonumber \\
&& +2 \sum_{i,j=1}^2 p_+^{(j)} p_-^{(i)} \; :  \psi_-^{(j)} \psi_+^{(i)}  V_{p}: (z) \bigg) 
 \; V_{-p} (w)\; T_{+1/2} (x)\; T_{-1/2} (y)} \nonumber \\
&=& \bcorr{\bigg(\pa_z V_p(z) + \sum_{k=1}^2 \; : i\left(p_{k} \; j_{k+d/2} - p_{k+d/2}\; j_{k}
\right)  \frac{1}{\sqrt{2}} \left( V^t_p - V^t_{-p} \right): (z) \nonumber \\
&& + p^2 \left( -\frac{1}{2} \; \frac{1}{z-x} + \frac{1}{2} \; \frac{1}{z-y} \right)
\;  V_{p} (z) \bigg) 
 \; V_{-p} (w)\; T_{+1/2} (x)\; T_{-1/2} (y)} .
\eeqn
The derivative of the vertex operator cancels with the corresponding term in the expression
with interchanged supercurrents in \mref{lang}. The second term in \mref{eins} vanishes when applying 
the U(1) currents $j_i$ to the vertex operators, using the OPE 
\beq
j_i (z)  \; V^t_{\pm p} (w) \sim  \pm \,  p_i (z-w)^{-1} V^t_{\pm p} (w)
\eeq
within the correlator.
Only the third term in \mref{eins} does not cancel with corresponding expressions, 
but U(1) charge conservation ensures that only the terms with  $i=j$ contribute.
Let $s_{\pm 1/2}^{(i)}$ signify the fermionic part of the twistfields $T_{\pm 1/2}$ with 
conformal dimension $h,\hb = 1/8$ in complex dimension $i$.
Then the  correlator leading to this third term in \mref{eins} is calculated in the relevant dimension 
according to
\beq
\lefteqn{\corr{ : \psi_-^{(i)}(w) \; \psi_+^{(i)}(z) : \; s_{+1/2}^{(i)} (x)\;  s_{-1/2}^{(i)}(y) } =} 
\nonumber \\
&=& \lim_{w\rightarrow z} \left[ \corr{ : \psi_-^{(i)}(w) \; \psi_+^{(i)}(z) : \; s_{+1/2}^{(i)}(x)
 \; s_{-1/2}^{(i)} (y) } - 
\frac{\corr{s_{+1/2}^{(i)} (x)\;  s_{-1/2}^{(i)}(y) }}
{w-z} \right] \nonumber \\
&=& \lim_{w\rightarrow z} \left[ (w-z)^{-1}\frac{ (w-y)^{1/2}(z-x)^{1/2}}{(w-x)^{1/2}(z-y)^{1/2}} 
(x-y)^{-1/4} 
- \frac{(x-y)^{-1/4}}{w-z} \right] \nonumber \\
&=& \left( -\frac{1}{2} \; \frac{1}{z-x} + \frac{1}{2} \; \frac{1}{z-y} \right)  
\corr{s_{+1/2}^{(i)} (x)\;  s_{-1/2}^{(i)}(y) } \; .
\eeq

Analogous calculations lead to the total result for the first and the second term in
\mref{lang}
\beqn
\label{erster-zweiter}
\lefteqn{\!\!\!\!\!\!\!\!\!\!\!\!\!\!\!\!\!\!\!\!\!\!\!\!\!
2 \; \left( -\frac{p^2}{2} \; \frac{1}{z-x} + \frac{p^2}{2} \; \frac{1}{z-y} 
-\frac{p^2}{2} \; \frac{1}{w-x} + \frac{p^2}{2} \; \frac{1}{w-y}\right) } \nonumber \\
&& \quad \quad \quad \quad \quad \times \; 
\bcorr{  V_{p} (z) \; V_{-p} (w)\; T_{+1/2} (x)\; T_{-1/2} (y)} \; .
\eeqn

Now, we turn to the third and fourth term in \mref{lang}. We can use the (anti)chiral properties
of  $T_{\pm 1/2}$ (i.e. $\ G^{\pm} (\eta) \;T_{\pm 1/2} (z)  \sim \; \mbox{finite}$) 
to add in terms with interchanged order of the contours, but equal sign.
Then, the usual contour argument (see e.g. \cite{Pol98}, \cite{Gab99}) together with
the OPE of the two supercurrents (see e.g. \cite{Gre96}) gives e.g.
\beqn
\label{zwei}
&& \!\!\!\!\!\!\!\!\!\!\!\!\!\!\!\!\!\!\!\!\!\!\! \bcorr{V_{p} (z)\; V_{-p} (w)\; 
\bigg( \Big( \oint_x \d \eta \; G^+(\eta) \; \oint_x \d \xi \; G^-(\xi) \nonumber \\
&&\!\!\!\!\!\!\!\!\!\!\!\!\!\!\!\!\!\! + \oint_x \d \xi \; G^-(\xi)\; \oint_x \d \eta \; G^+(\eta) \Big) \; 
T_{+1/2} (x) \bigg)\; T_{-1/2} (y)} \nonumber \\
&=& \bcorr{V_{p} (z)\; V_{-p} (w)\; \left( \oint_x \d \xi  \; \oint_{\xi} \d \eta \; G^+(\eta) G^-(\xi)
T_{+1/2} (x) \right)\; T_{-1/2} (y)} \nonumber \\
&=&  \bcorr{V_{p} (z)\; V_{-p} (w)\; \left( \oint_x \d \xi  \; (2\; \Tc(\xi) + \pa_{\xi} J (\xi))
T_{+1/2} (x) \right)\; T_{-1/2} (y)} \nonumber \\
&=& 2\;  \pa_x \; \bcorr{V_{p} (z)\; V_{-p} (w)\; T_{+1/2} (x) \; T_{-1/2} (y)} \; ,
\eeqn
where $\Tc(\xi)$ and $J (\xi)$ signify the Virasoro field and the U(1) current, respectively, 
of the N=2 superconformal algebra.
The last step uses the conformal algebra of primary fields (e.g. \cite{Gre96}). Calculating the derivative
of the correlator \mref{verkorr2} given in the appendix
\beq
\lefteqn{\!\!\!\!\!\! \pa_x \bcorr{V^t_{p} (z)\; V^t_{\mp p} (w)\; T_{+1/2} (x) \; T_{-1/2} (y)} } \nonumber \\
&=& \pa_x \bigg[4^{-p^2} (x-y)^{-1+(p\mp p)^2/2} ((z-x)(z-y)(w-x)(w-y))^{-p^2/2}
\nonumber \\
&& \times \quad \left(\frac{((z-x)^{1/2}(w-y)^{1/2}+(z-y)^{1/2}(w-x)^{1/2})^2}
{(w-z)}  \right)^{\pm p^2} \bigg] \nonumber \\
&=& \bigg( \frac{\frac{(p\mp p)^2}{2}-1}{x-y} + 
\frac{p^2}{2} \; \frac{1}{z-x} + \frac{p^2}{2} \; \frac{1}{w-x} \nonumber \\
&&\mp p^2 \; \frac{(z-x)^{-1/2} (w-y)^{1/2} + (z-y)^{1/2} (w-x)^{-1/2}}
{(z-x)^{1/2} (w-y)^{1/2} + (z-y)^{1/2} (w-x)^{1/2}} \bigg) \nonumber \\
&& \times   \bcorr{V^t_{p} (z)\; V^t_{\mp p} (w)\; T_{+1/2} (x) \; T_{-1/2} (y)} \; .
\eeq
we get the total contribution of the third and fourth terms in \mref{lang}
\beqn
\label{dritter-vierter}
\lefteqn{2\; ( \pa_x - \pa_y) \; \bcorr{V_{p} (z)\; V_{-p} (w)\; T_{+1/2} (x) \; T_{-1/2} (y)} =}\nonumber \\
&=& 2 \; \bigg( 2 \; \frac{p^2-1}{x-y} 
+ \frac{p^2}{2} \left( \frac{1}{z-x} + \frac{1}{w-x} - \frac{1}{z-y} - \frac{1}{w-y} \right) 
\bigg)\nonumber \\
&& \times  \bcorr{V_{p} (z)\; V_{-p} (w)\; T_{+1/2} (x) \; T_{-1/2} (y)} \nonumber \\
&+& 2 \; \bigg(p^2 (x-y)^{-1} (z-x)^{-1/2} (z-y)^{-1/2}  (w-x)^{-1/2}  (w-y)^{-1/2} 
\Big( (z-x) (w-y)  \nonumber \\
&&+ (z-y) (w-x) \Big) 
+ p^2 (z-x)^{-1/2}  (z-y)^{-1/2}  (w-x)^{-1/2}  (w-y)^{-1/2} (x-y) \bigg) \nonumber \\
&&  \times   \bcorr{\frac{1}{\sqrt{2}} \Big(V^t_{p} - V^t_{-p}\Big) (z)\; 
\frac{1}{\sqrt{2}} \Big(V^t_{p} - V^t_{-p}\Big) (w) \; T_{+1/2} (x) \; T_{-1/2} (y)} \; .
\eeqn

Now, only the last two terms in \mref{lang} are still left. Again, using \mref{super} we get
for the fifth term in \mref{lang}
\beq
\lefteqn{ \bcorr {\left(\oint_z \d \xi \; G^-(\xi) V_{p} (z)\right) 
\left( \oint_w \d \eta \; G^+(\eta) \; V_{-p} (w)\;\right)  \; T_{+1/2} (x)\; T_{-1/2} (y)} =} \nonumber \\
&=& \bcorr{ \bigg(\sum_{i=1}^2 \sqrt{2} \left(\psi_-^{(i)} (z) p_+^{(i)} \right) \frac{1}{\sqrt{2}}
\left( V_p^t-V_{-p}^t \right) (z) \bigg) \nonumber \\
&& \bigg(\sum_{j=1}^2 \sqrt{2} \left(\psi_+^{(j)} (w) p_-^{(j)} \right) \frac{1}{\sqrt{2}}
\left( V_p^t-V_{-p}^t \right) (w) \bigg) \; T_{+1/2} (x) \; T_{-1/2} (y)}\nonumber \\
&=&  2\;\bigg(\sum_{i,j=1}^2 p_+^{(i)} p_-^{(j)} \corr{ \psi_-^{(i)} (z)  \psi_+^{(j)} (w)
\prod_{b=1}^2 s^{(b)}_{+1/2} (x) s^{(b)}_{-1/2} (y) } \bigg)
\bigg(\corr{\prod_{b=1}^2 s^{(b)}_{+1/2} (x) s^{(b)}_{-1/2} (y) } \bigg)^{\!\!\! -1} \; \nonumber \\
&& \times \bcorr{\frac{1}{\sqrt{2}} \Big(V^t_{p} - V^t_{-p}\Big) (z)\; 
\frac{1}{\sqrt{2}} \Big(V^t_{p} - V^t_{-p}\Big) (w) \; T_{+1/2} (x) \; T_{-1/2} (y)} \nonumber \\
&=& p^2 (z-w)^{-1} (z-x)^{-1/2}  (z-y)^{1/2}  (w-x)^{1/2}  (w-y)^{-1/2} \nonumber \\
&& \times  \bcorr{\frac{1}{\sqrt{2}} \Big(V^t_{p} - V^t_{-p}\Big) (z)\; 
\frac{1}{\sqrt{2}} \Big(V^t_{p} - V^t_{-p}\Big) (w) \; T_{+1/2} (x) \; T_{-1/2} (y)} \; .
\eeq
Adding in the similar contribution of the sixth term of \mref{lang}, the last two terms together
give
\beqn
\label{fuenfter-sechster}
&&\!\!\!\!\!\!\bcorr{\left(\oint_z \d \xi \; G^-(\xi) V_{p} (z)\right) 
\left( \oint_w \d \eta \; G^+(\eta) \; V_{-p} (w)\;\right)  \; T_{+1/2} (x)\; T_{-1/2} (y)} \nonumber \\
&& - \;  \bcorr {\left(\oint_z \d \eta \; G^+(\eta) V_{p} (z)\right) 
\left( \oint_w \d \xi \; G^-(\xi) \; V_{-p} (w)\;\right)  \; T_{+1/2} (x)\; T_{-1/2} (y)} \nonumber \\
&=& - 2\; p^2 (z-x)^{-1/2}  (z-y)^{-1/2}  (w-x)^{-1/2}  (w-y)^{-1/2} (x-y)  \nonumber \\
&&  \times  \bcorr{\frac{1}{\sqrt{2}} \Big(V^t_{p} - V^t_{-p}\Big) (z)\; 
\frac{1}{\sqrt{2}} \Big(V^t_{p} - V^t_{-p}\Big) (w) \; T_{+1/2} (x) \; T_{-1/2} (y)} \; .
\eeqn

Adding the results of \mref{erster-zweiter}, \mref{dritter-vierter} and 
\mref{fuenfter-sechster}, the total result reads
\beqn
\label{erg}
\tilde{Z} &= &  2 \; \frac{p^2-1}{x-y} \;
\bcorr{V_{p} (z)\; V_{-p} (w)\; T_{+1/2} (x) \; T_{-1/2} (y)} \nonumber \\
&+& p^2 (x-y)^{-1} (z-x)^{-1/2}  (z-y)^{-1/2}  (w-x)^{-1/2}  (w-y)^{-1/2} \nonumber \\
&& \times \Big( (z-x) (w-y) + (z-y) (w-x) \Big) \nonumber \\   
&& \times \bcorr{\frac{1}{\sqrt{2}} \Big(V^t_{p} - V^t_{-p}\Big) (z)\; 
\frac{1}{\sqrt{2}} \Big(V^t_{p} - V^t_{-p}\Big) (w)\; T_{+1/2} (x) \; T_{-1/2} (y)} \; .
\eeqn

As already remarked earlier, the calculation for the antiholomorphic part runs independently
and in exactly the same fashion. Call $p_l$ the U(1) charge w.r.t.\ the left moving, i.e.\ holomorphic 
U(1) current, $p_r$ the U(1) charge for the right moving, i.e.\ antiholomorphic U(1) current.
Hence, the total result for  \mref{eigentlich} is given 
by
\beqn
\label{eigentlich2}
Z &= & \bigg[2\; \frac{p_l^2-1}{x-y} \; - p_l^2 (x-y)^{-1} (z-x)^{-1/2}  (z-y)^{-1/2}  
(w-x)^{-1/2}  (w-y)^{-1/2} \nonumber \\
&& \Big( (z-x) (w-y) + (z-y) (w-x) \Big)\bigg] \nonumber \\ 
&& \times \bigg[2\; \frac{p_r^2-1}{\xb -\yb} \; - p_r^2 (\xb -\yb )^{-1} (\zb -\xb )^{-1/2}  (\zb -\yb )^{-1/2}  
(\wb -\xb )^{-1/2}  (\wb -\yb)^{-1/2} \nonumber \\
&& \Big( (\zb -\xb ) (\wb -\yb ) + (\zb -\yb ) (\wb -\xb) \Big)\bigg] \nonumber \\
&& \times \bcorr{V_{p}^t (z,\zb)\; V_{-p}^t (w,\wb)\; T_{+1/2} (x,\xb) \; T_{-1/2} (y,\yb)} \nonumber \\
&+&  \bigg[2\;\frac{p_l^2-1}{x-y} \; + p_l^2 (x-y)^{-1} (z-x)^{-1/2}  (z-y)^{-1/2}  
(w-x)^{-1/2}  (w-y)^{-1/2} \nonumber \\ 
&& \Big( (z-x) (w-y) + (z-y) (w-x) \Big)\bigg] \nonumber \\ 
&& \times \bigg[2\; \frac{p_r^2-1}{\xb -\yb } \; + p_r^2 (\xb -\yb )^{-1} (\zb -\xb)^{-1/2}  (\zb -\yb)^{-1/2}  
(\wb -\xb )^{-1/2}  (\wb -\yb )^{-1/2} \nonumber \\
&& \Big( (\zb -\xb) (\wb -\yb) + (\zb -\yb ) (\wb -\xb ) \Big)\bigg] \nonumber \\
&& \times \bcorr{V_{p}^t (z,\zb)\; V_{p}^t (w,\wb)\; T_{+1/2} (x,\xb) \; T_{-1/2} (y,\yb)} \; .
\eeqn

For the calculation of the perturbation of the two-vertex-correlator, we now want
to specialise to the case of small $p_l^2,p_r^2 < 1$. The vector of the left moving and right moving
U(1) charges can be shown to lie on an even integer charge lattice in the underlying toroidal theory, 
especially $p_l^2-p_r^2 \in 2\,\Z$ (see e.g.\ \cite{Pol98}). As a direct consequence of the above
assumption, we therefore deduce $p_l^2 = p_r^2 \equiv p^2$.
The perturbation in second order, according to \mref{direkt2},
\mref{eigentlich} and \mref{eigentlich2}, simplifies to
\beqn
\label{int}
\int \d^2x  \int \d^2y\; Z =
 (z-w)^{-p_l^2} \; (\zb-\wb)^{-p_r^2} \;  \int \d^2x \; 
\frac{\bet{z-w}^2}{\bet{x-z}^2 \; \bet{x-w}^2} \; (V_1 + V_2) \; ,
\eeqn
where $V_1$ corresponds to the first term in the sum in \mref{eigentlich2},
$V_2$ to the second, and where the integrals $V_i$ are given by
\beq
V_1 & \equiv & 4^{-p_l^2-p_r^2} \;\int d^2\zeta \; \frac{1}{\bet{1-\zeta}^4} \; 
\Big( 2 \;(p_l^2-1) -p_l^2 \; \zeta^{-1/2} \; 
(1+\zeta) \Big) \nonumber \\ 
&& \Big( 2 \;(p_r^2-1) -p_r^2 \; \zetab^{-1/2} \; 
(1+\zetab) \Big) \; \zeta^{-p_l^2/2} \; \zetab^{-p_r^2/2} \; (1-\zeta)^{p_l^2} \; (1-\zetab)^{p_r^2}
\nonumber \\
&& \bigg( \frac{1+\zeta^{1/2}}{1-\zeta^{1/2}} \bigg)^{p_l^2} \;  
\bigg( \frac{1+\zetab^{1/2}}{1-\zetab^{1/2}} \bigg)^{p_r^2} \nonumber \\
V_2 & \equiv & 4^{-p_l^2-p_r^2} \;\int d^2\zeta \; \frac{1}{\bet{1-\zeta}^4} \; 
\Big( 2\; (p_l^2-1) +p_l^2 \; \zeta^{-1/2} \; 
(1+\zeta) \Big) \nonumber \\ 
&& \Big( 2 \;(p_r^2-1) +p_r^2 \; \zetab^{-1/2} \; 
(1+\zetab) \Big) \; \zeta^{-p_l^2/2} \; \zetab^{-p_r^2/2} \; (1-\zeta)^{p_l^2} \; (1-\zetab)^{p_r^2}
\nonumber \\
&& \bigg( \frac{1-\zeta^{1/2}}{1+\zeta^{1/2}} \bigg)^{p_l^2} \;  
\bigg( \frac{1-\zetab^{1/2}}{1+\zetab^{1/2}} \bigg)^{p_r^2} \; ;
\eeq
the four point correlator calculated in the appendix has been plugged in explicitely.
The simplification in \mref{int} uses the SL(2,$\C$) transformation
\beq
\zeta(\eta) = \frac{\eta-z}{\eta-w} \; \frac{x-w}{x-z} \; ,
\eeq
which maps $z \mapsto 0$, $w\mapsto w' \rightarrow \infty$, $x\mapsto 1$ 
and $y \mapsto \zeta$.

Both integrals $V_i$ can be merged to a single integral over the two sheeted Riemann surface
of $t \equiv \pm \sqrt{\zeta}$. However, this presentation would make the following
discussion of singularities considerably harder.


\section{Singularities\label{sing}}

As explained below equation \mref{direkt2}, we have to identify and subtract the disconnected part 
of the integrand of \mref{int} and suitably regulate the remaining integral.
There are two sources of singularities in  \mref{int} for the case of small $p^2 \equiv p_l^2=p_r^2$.
First, for $z \rightarrow w$ we have a mixing of the two vertex operators  
$V^t_p (z,\zb)$ and  $V^t_{-p}(w,\wb)$ with the identity; furthermore, for $x \rightarrow y$ the
two twistfields $S_{+1/2}(x,\xb)$ and  $S_{-1/2}(y,\yb)$ also mix with the
identity. This channel is responsible for the disconnected part of \mref{int} which 
produces the singularity  $\bet{1-\zeta}^{-4}$ in  $V_1$. 
Second, for  $z \rightarrow w$ the two vertex operators $V^t_p (z,\zb)$ and  $V^t_{p}(w,\wb)$
mix with the vertex operator $V^t_{-2p} (z,\zb)$; the same mixing occurs for the two twistfields
$S_{+1/2}(x,\xb)$ and $S_{-1/2}(y,\yb)$ for $x \rightarrow y$. This channel is responsible
for the  $\zeta \rightarrow 1$ singularity in $V_2$.
To understand this seemingly unusual mixing of $S_{+1/2}(x,\xb)$ and $S_{-1/2}(y,\yb)$ with
the vertex operator  $V^t_{-2p} (z,\zb)$, we have a closer look at the orbifold
construction itself. In the twisted sector, the momenta  $p$ and $-p$ are identified 
by the action of the orbifold group and
exactly this fact guarantees that the conservation of momentum is not violated in the above OPE.

These singularities call for a proper regularisation procedure. In position space, 
this regularisation is best achieved by introducing a cut-off $\eps \rightarrow 0$ around the
singularities and by subtracting the singular parts in $\eps$. But this can be shown
to be equivalent to the following much more practicable procedure. First, one subtracts
the terms that cause these singularities from the integrand. Then, one can integrate
this integrand over the whole complex plane and acquires the correct result. (By definition, a
term which causes a singularity has to behave exactly as the original integrand close to the
singularity, and the integration of this singular term over the whole plane, cut off with the same
$\epsilon$ at the singularity, should only exhibit an $\epsilon$ dependence. This is hence the same
$\epsilon$ dependence as the one of the integral of the original integrand cut off close to the same 
singularity.)  

Now, we want to write down the precise form of these two singular terms which are to be
subtracted.
Using the correlator of two ground state twistfields
\beq
\corr{ T_{+1/2} (x,\xb)\; T_{-1/2} (y,\yb)} =  \bet{x-y}^{-2} \; ,
\eeq
an analogous turning around of the contours as in the case of the full four point function
\mref{vertex1} leads to the correlator of the superpartners
\beq
\corr{ S_{+1/2} (x,\xb)\; S_{-1/2} (y,\yb)} =  4 \; \bet{x-y}^{-4} \; .
\eeq
Hence, the whole disconnected contribution to $V_1$ reads
\beq
\corr{V_{p}^t (z,\zb)\; V_{-p}^t (w,\wb)}\; \corr{ S_{+1/2} (x,\xb)\; S_{-1/2} (y,\yb)}
= 4 \; (z-w)^{-p_l^2} \; (\zb-\wb)^{-p_r^2} \; \bet{x-y}^{-4} \; .
\eeq
Doing an  SL(2,$\C$) transformation as above, we get our final result for this singular part
of the integral 
\beq
(z-w)^{-p_l^2} \; (\zb-\wb)^{-p_r^2} \;  \int \d^2x \; 
\frac{\bet{z-w}^2}{\bet{x-z}^2 \; \bet{x-w}^2} \; V_1^S
\eeq
with
\beq
V_1^S \equiv 4\; \int \d^2\zeta \frac{1}{\bet{1-\zeta}^4} \; .
\eeq
$V_1^S$ is exactly the $\zeta \rightarrow 1$ singularity of $V_1$. But as we only want to integrate
over the connected part of the total integrand in \mref{int}, we have to replace $V_1$ by $V_1 - V_1^S$;
this does not contain any singularity any more.

The second singularity, the singular part of  $V_2$, is given for  $p^2 \le \frac{1}{2}$ by
\beqn
\label{fuehr}
V_2^S \equiv 16^{-p_l^2-p_r^2} \; 4 \;  (2p_l^2 -1) \; (2p_r^2 -1)   \; \int \d^2\zeta \;
(1-\zeta)^{2p_l^2-2} \;  (1-\zetab)^{2p_r^2-2} \; .
\eeqn
For $\frac{1}{2}<p^2<1$,  $V_2$ is even finite. 
We now want to show that this singularity really originates in the mixing
with vertex operators of charge $\pm 2p$.
First, we notice that the correlator of three vertex operators is given by conformal invariance
\beqn
\label{teil1}
\lefteqn{\corr{V_{p}^t (z,\zb)\; V_{p}^t (w,\wb) \;V_{-2p}^t (\xi,\bxi)} =} \nonumber \\
&=& (z-w)^{p_l^2} \; (\zb-\wb)^{p_r^2} \;(z-\xi)^{-2p_l^2} \; (\zb-\bxi)^{-2p_r^2} 
\;(w-\xi)^{-2p_l^2} \; (\wb-\bxi)^{-2p_r^2} \; .
\eeqn
Analogously, conformal invariance fixes the correlator of two groundstate twistfields
with a vertex operator
\beq
\lefteqn{\corr{V_{2p}^t (\xi,\bxi) \; T_{+1/2} (x,\xb)\; T_{-1/2} (y,\yb)} =} \nonumber \\
&=& C \; (x-y)^{2\,p_l^2  -1} \;  (\xb-\yb)^{2\,p_r^2  -1} 
\;(x-\xi)^{-2\,p_l^2}  \;(\xb-\bxi)^{-2\,p_r^2} 
\;(y-\xi)^{-2\,p_l^2}  \;(\yb-\bxi)^{-2\,p_r^2} \; 
\eeq
and a similar calculation as for the full four point function leads to
the corresponding result with supertwistfields
%
%
\beqn
\label{teil2}
\lefteqn{\corr{V_{2p}^t (\xi) \; S_{+1/2} (x)\; S_{-1/2} (y)} =} \nonumber \\
&=& 4 \; C \; (2p_l^2 -1) \; (2p_r^2 -1) \; (x-y)^{2\,p_l^2  -2} \;  (\xb-\yb)^{2\,p_r^2  -2} 
\;(x-\xi)^{-2\,p_l^2}  \nonumber \\
&& \;(\xb-\bxi)^{-2\,p_r^2} 
\;(y-\xi)^{-2\,p_l^2}  \;(\yb-\bxi)^{-2\,p_r^2} \; .
\eeqn
Even without doing the integration over $\xi$, one can identify the leading singularity
in the product of \mref{teil1} and \mref{teil2} with the singular part of $V_2$ \mref{fuehr}.
This identification fixes the normalisation factor to $C=16^{-p_l^2-p_r^2}$. Further
terms in this product of \mref{teil1} and \mref{teil2} give higher orders of $(x-y)$ for 
$x \rightarrow y$ and hence do not contribute to the singular part of the integral.
The total regularisation for $V_2$ reads $V_2 - V_2^S$ in the case $p^2\le \frac{1}{2}$;
for  $\frac{1}{2}<p^2<1$ no regularisation is necessary.


As these two singularities are the only ones for small $p_l^2$, $p_r^2$, the total regularisation
of $V_1+V_2$ in \mref{int} is given by 
\beqn
\label{num}
V &\equiv& (V_1 - V_1^S) + (V_2 - V_2^S)  \quad \quad \quad \mbox{for} \;  p^2\le\frac{1}{2} \nonumber \\
V &\equiv& (V_1 - V_1^S) + V_2 \quad \quad \quad \quad \quad \mbox{for} \; \frac{1}{2}<p^2<1 \; .
\eeqn

For larger $p_l^2 \ge 1$ or  $p_r^2 \ge 1$, we encounter further singularities also for 
$\zeta \rightarrow 0$ and $\zeta \rightarrow \infty$. These are the result of a mixing
of $V_p (w,\wb)$ and $S_{\pm} (x,\xb)$ with ground state twistfields (see e.g. \cite{HV87}
for a discussion of such processes on string theoretic grounds).


\section{Results}

We now want to calculate the perturbation \mref{int} subtracting the singularities as
discussed in section \ref{sing}. In order to perform the $x$-integral in \mref{int}, we use
the translational invariance of correlation functions to set $w=0$ and apply the
SL(2,$\C$) transformation
\beq
\gamma : \quad \zeta'(\zeta) = \frac{z\zeta}{z-\zeta} \quad \quad \quad \quad
\gamma^{-1} : \quad \zeta(\zeta') = \frac{z\zeta'}{\zeta'+z}
\eeq
which maps $0$ to $0$ and $z$ to $z'\mapsto \infty$. Introducing a proper cut-off $\eps$,
the logarithmically divergent $x$-integral in \mref{int} can be calculated to
\beq
\int_{\bet{x-z} > \eps \atop \bet{x} > \eps}
 \d^2x \; \frac{\bet{z}^2}{\bet{x-z}^2 \; \bet{x}^2} = 2 \pi \log \frac{\bet{z}^2}{\eps^2} \; .
\eeq
As shown in \mref{direkt2}, these logarithmic divergencies correspond to
the variation of the conformal dimension of the vertex operators $V_p (z,\zb)$.

\TABULAR{c | c || c | c}{
$p^2$ & V & $p^2$ & V \\ \hline
0.00000   &      0.00000  & 0.26000    &     -0.71670 \\ 
0.01000   &      -0.06111 & 0.272211   &     -0.71256 \\
0.01562   &      -0.09401 & 0.277008   &     -0.71007 \\
0.02250   &      -0.13281 & 0.280277   &     -0.70810 \\ 
0.04000   &      -0.22487 & 0.284444   &     -0.70525 \\
0.06250   &      -0.32982 & 0.294117   &     -0.69722 \\
0.07840   &      -0.39543 & 0.302500   &     -0.68865 \\
0.09000   &      -0.43903 & 0.30864    &     -0.68141 \\ 
0.11111   &      -0.50954 & 0.36000    &     -0.58841 \\ 
0.16000   &      -0.63164 & 0.40111    &     -0.47008 \\ 
0.20250   &      -0.69395 & 0.44444    &     -0.29899 \\ 
0.22145   &      -0.70917 & 0.49000    &     -0.06109 \\ 
0.22438   &      -0.71084 & 0.50000    &     0.00000 \\
0.22873   &      -0.71299 & 0.56250    &     0.46758 \\
0.23529   &      -0.71548 & 0.64000    &     1.31499 \\
0.24000   &      -0.71670 & 0.69444    &     2.17848 \\
0.25000   &      -0.71775 & &
}{Numerical results\label{tab}}

Hence, the whole information about the deformation is given in the integral $V$ as in
\mref{num}. We solved the integral numerically for various $p^2<1$ using the program Mathematica.
However, although the integral converges there are still divergencies in the integrand
which have to be taken care of. The regularised integral  $V_1$ still contains divergencies
of the form $(1-\zeta)^{-2}$ or $(1-\zetab)^{-2}$ which we handle by cutting out
an $\eps$-ball around $\zeta=1$ of size $\eps=0.01$. Integration over 
the first terms of a Taylor expansion around  $\zeta=1$ 
gives the approximation that the contribution of this ball to the total integral amounts
to about $10^{-6}$.
In order to handle the  $\bet{1-\zeta}^{-r}$, $r<2$, singularities in the regularised $V_2$, 
we subtract the expression 
\beq
\tilde{V}^S_2 &\equiv& 16^{-2p_l^2} \; 4 \;  (2p^2 -1)^2 \\
&& \qquad \int \d^2\zeta \; 
\bet{1-\zeta}^{4p_l^2-4} \; \bet{\zeta}^{-2p^2}
\eeq
instead of $V_2^S$. $V_2 - \tilde{V}^S_2$ behaves much better numerically and the difference
$\tilde{V}^S_2 - V_2^S$ is known as {\em Virasoro Shapiro amplitude} in the literature
(see e.g.  \cite{Pol98})
\beqn
\label{shap}
\tilde{V}^S_2 - V_2^S &=& 16^{-2p^2}  \; 4 \;  (2p^2 -1)^2 \;\int \d^2\zeta \; 
\bet{1-\zeta}^{4p^2-4} \left( \bet{\zeta}^{-2p^2}  -1 \right) \nonumber \\
&=& \pi \frac{\Gamma(1-p^2)^2 \; \Gamma(2p^2-1)}{\Gamma(p^2)^2 \; \Gamma(2-2p^2)} \; .
\eeqn
$V_2^S$ has to be regarded as regularisation of $\tilde{V}^S_2$ using the regularisation
procedure of a cut-off $\eps$ again. For $0<p^2 < \frac{1}{2}$ we get the Virasoro Shapiro amplitude
by analytical continuation, for the case $\frac{1}{2} < p^2 < 1$ 
the regularisation term 
$V_2^S$ vanishes, $\tilde{V}^S_2$ does not diverge and directly equals the finite
Virasoro Shapiro amplitude (see e.g.\cite{Col84} for these regularisation procedures).
The case $p^2=0$ is analytically obvious, for $p^2=1/2$ the regularisation term for $V_2$
vanishes anyhow.

\EPSFIGURE{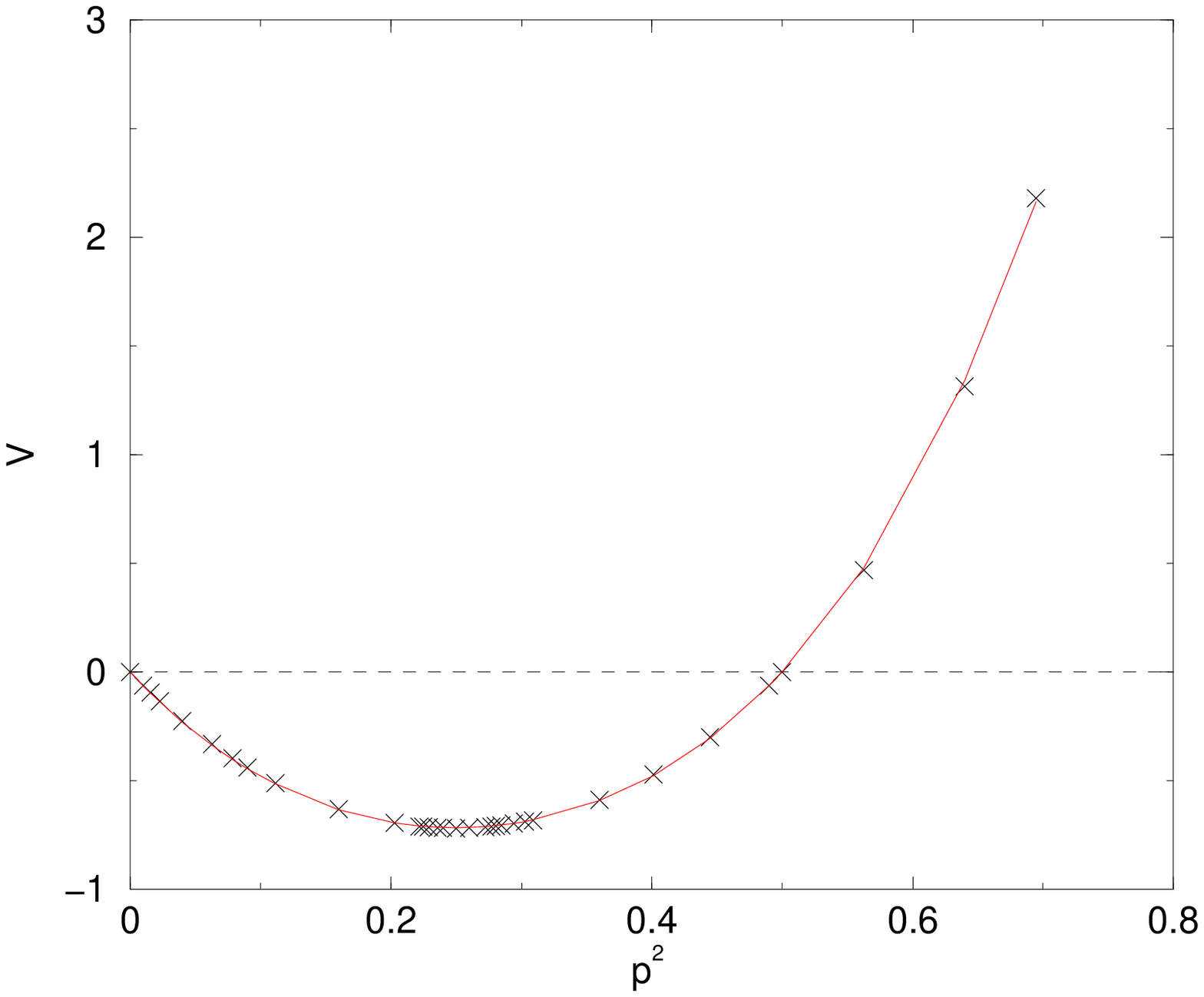,width=10cm}{Graphical representation of the results\label{plot}}
The results of this numerical procedure are depicted in table \ref{tab} for various $p^2$.
A plot of these results is given in figure \ref{plot}. There are several things to
notice in this plot. The reflection symmetry around the axis $p^2=1/4$ seems to be most obvious.
This strongly suggests an even function in $(p^2-1/4)$. Furthermore, the
function seems to acquire its minimum at $p^2=1/4$ and the perturbation seems to vanish at
$p^2=1/2$. This second observation is another strong indication for the reflection symmetry
around the axis $p^2=1/4$ as the perturbation vanishes for $p^2=0$ analytically.

These observations lead us to attempt the following fit (depicted in figure \ref{plot}) 
\beq
V &=& -0.71775+a_2 \; \left(p^2-\frac{1}{4}\right)^2 \,
+ \, a_4 \; \left(p^2-\frac{1}{4}\right)^4 \; .
\eeq
This fit produces values of $a_2=10.045 \pm 0.054$ and 
$a_4=22.940\pm 0.363$. These errors are given by the asymptotic covariance matrix. Especially
in the range $0<p^2 < \frac{1}{2}$ this fit seems to be quite accurate; for $p^2> 1/2$ contributions
of higher orders in $\left(p^2-\frac{1}{4} \right)^2$ seem to become more important.



\section{Geometrical interpretation}

The calculations in this article have been performed on the $\Z_2$ orbifold subvariety
of K3 moduli space. In contrast to deformations with U(1) currents which describe the relations
between theories on the $\Z_2$ orbifold subvariety we have calculated deformations
with twistfields that deform away from this subvariety. In the geometrical picture
a deformation with twistfields leads to a transition from singular orbifolds to smooth
manifolds. This transition is achieved by the so-called {\em blowing-up} of the 
orbifold fixed point singularities \cite{DHVW85}, \cite{Asp94}, \cite{Asp96}. The
blowing-up of a specific fixed point in the geometrical picture corresponds to
the deformation with the twistfield located at exactly this fixed point.

Blown-up orbifolds have already been studied in the context of the Landau-Ginzburg description
by giving the blowing-up modes (i.e.\ the twistfields) a non-vanishing vacuum expectation
value \cite{Cve87}. A strictly conformal field theoretic approach, however, is not known
to the author.


\section{Conclusion and outlook}

In this article, we have calculated the deformation of the conformal dimension of vertex operators
deforming the $\Z_2$ orbifold model with twistfields in second order perturbation theory. 
This was a first attempt to this kind of problem and, hence, there are still a lot of
open questions to be answered. First, we hope to validate analytically some of the conjectures
based on the numerical results in this article in due course. The cases of $p^2>1$ and of general
$\Z_N$ orbifolds should be analysable by the same methods, at least to second order in
perturbation theory. Higher orders, however, seem to become more difficult very quickly. 
And, of course, an attempt should be made to integrate these deformations to get hold on
further subvarieties in moduli space. At least, this should be possible for directions in
moduli space showing a high degree of symmetry.


\vspace{0.7cm}
\noindent
{\bf Acknowledgements.}
I would like to thank my supervisor W. Nahm for suggesting this exciting topic and
for many very helpful discussions. I am also grateful to K. Wendland and A. Wi\ss kirchen
for discussions and to M. R\"osgen for remarks on the manuscript.


\begin{appendix}

\section{The four point correlator of two ground state twistfields with two vertex operators
\label{verkorr}}

The four point correlator of two ground state twistfields with two vertex operators 
in the $\Z_2$ orbifold has
already been calculated in \cite{HV87} based on results about off-shell string amplitudes
in \cite{SW74}, \cite{CF75} and \cite{Gre76}, but neglecting the correct
zero mode contribution. However, this has then been discussed in \cite{EJLM92} and \cite{Jun92}.
As this correlator is the basis for our calculations, we want to present a short
derivation just using basic properties of conformal field theory.

To calculate the correlation function
\beqn
\label{ver-korr}
\corr{V^t_{p_1} (w,\wb)\; V^t_{p_2} (v,\vb)\; T_{1/2} (x,\xb)\; T_{-1/2} (y,\yb) }
\eeqn
we want to use the OPE of the vertex operators $V^t_{p} (z,\zb)$ with any U(1) current $j(z) = i\pa X$
of the underlying toroidal theory.
In order to derive this OPE, we use the mode expansion of $j(z)$
\beq
j(z) = \sum_{n=0}^{\infty} j_n z^{n-1} \quad \quad \quad j_{n} \cet{\phi} = 0 \quad \forall \; n<0\; 
\eeq
and the corresponding mode expansion of the bosonic part of the holomorphic Virasoro field 
\beq
{\cal T}_{bosonic}= \sum_{i=1}^{d} :j_i \;j_i:(z) \; .
\eeq
Therefore, applying the Virasoro modes to the state $\cet{V^t_p}$, which corresponds to the
vertex operator $V^t_p (0,0)$, the mode expansions give
\beq
\cet{\pa V^t_p} &=& L_1 \cet{V^t_p} \nonumber \\
&=& \frac{1}{2} (j_{0\, a}\; j_1^a+j_{1\, a} \; j_0^a) \cet{V^t_p} \nonumber \\
&=& p_aj_1^a \cet{V^t_p} \; .
\eeq
The summations run over all d holomorphic currents in a real--d dimensional theory.
This leads to the OPE
\beqn
\label{auch}
p_aj^a(z) \; V^t_p (w) = \frac{p^2}{z-w} \; V^t_p (w) + \pa_w V^t_p (w) + O (z-w) \; .
\eeqn
Also define the scalar product $p_{1\, a} \; p_2^a = p_1.p_2$, $p_1^2=p_1.p_1$ for the charge vectors
w.r.t.\ to the holomorphic U(1) currents.
Our goal is to find a first order differential equation for \mref{ver-korr}.
To achieve this we turn to the following five point correlator
\beqn
\label{fuenf}
\lefteqn{\!\!\!\!\!\!\!\!\!\!\!\!\!\!\!\sqrt{(z-x)(z-y)} \;
\corr{p_{1a} j^a(z) \; V^t_{p_1} (w,\wb)\; V^t_{p_2} (v,\vb)\; T_{1/2} (x,\xb)\; T_{-1/2} (y,\yb) }
=} \nonumber \\
&=& \left(\sqrt{(w-x)(w-y)} \;\frac{p_1^2}{z-w} + \sqrt{(v-x)(v-y)} \;\frac{p_1.p_2}{z-v} \right)
\nonumber \\
&& \times \corr{V^t_{p_1} (w,\wb)\; V^t_{p_2} (v,\vb)\; T_{1/2} (x,\xb)\; T_{-1/2} (y,\yb) } \; .
\eeqn
This identity is justified by noticing that
the square root on the left hand side of \mref{fuenf} just cancels the branching behaviour
of the current $j(z)$ around the twistfields $T_{\pm1/2}$. Hence, 
we have a meromorphic function in $z$ on the right hand side. The meromorphic behaviour of this
right hand side is completely determined by the singular behaviour of the OPE
between the current and the two vertex operators and the fact that every
U(1) current drops off as $1/z^2$ at infinity. The behaviour with respect to the
other coordinates is already contained in the four point correlator \mref{ver-korr}. 

Now we are ready to calculate the derivatives of \mref{ver-korr} with respect to $w$ and $v$
using the above OPE \mref{auch} and the identity \mref{fuenf}
\beq
\lefteqn{ 
\pa_w \corr{V^t_{p_1} (w,\wb)\; V^t_{p_2} (v,\vb)\; T_{1/2} (x,\xb)\; T_{-1/2} (y,\yb) } =} \nonumber \\
&=& \lim_{z\rightarrow w} 
\bigg[ \corr{p_{1a} j^a(z) \; V^t_{p_1} (w,\wb)\; V^t_{p_2} (v,\vb)\; T_{1/2} (x,\xb)\; T_{-1/2} (y,\yb) }
\nonumber  \\
&&  - \frac{p_1^2}{z-w} \corr{V^t_{p_1} (w,\wb)\; V^t_{p_2} (v,\vb)\; T_{1/2} (x,\xb)\; T_{-1/2} (y,\yb) }  
\bigg] \nonumber \\
&=& \lim_{z\rightarrow w} \left[ \left( \frac{(w-x)^{1/2}(w-y)^{1/2}}{(z-x)^{1/2}(z-y)^{1/2}} \;
\frac{p_1^2}{z-w} + \frac{(v-x)^{1/2}(v-y)^{1/2}}{(z-x)^{1/2}(z-y)^{1/2}} \;
\frac{p_1p_2}{z-v} \right) - \frac{p_1^2}{z-w} \right] \nonumber \\ 
&& \times \corr{V^t_{p_1} (w,\wb)\; V^t_{p_2} (v,\vb)\; T_{1/2} (x,\xb)\; T_{-1/2} (y,\yb) } \nonumber \\
&=&\left[\frac{-p_1^2}{2} \left( \frac{1}{w-x} + \frac{1}{w-y} \right) + \frac{ p_1p_2}{w-v} 
\frac{(v-x)^{1/2}(v-y)^{1/2}}{(w-x)^{1/2}(w-y)^{1/2}} \right] \nonumber \\
&& \times \corr{V^t_{p_1} (w,\wb)\; V^t_{p_2} (v,\vb)\; T_{1/2} (x,\xb)\; T_{-1/2} (y,\yb) }  \; .
\eeq
This leads to the logarithmic derivative
\beq
\lefteqn{\pa_w \log \corr{V^t_{p_1} (w,\wb)\; V^t_{p_2} (v,\vb)\; T_{1/2} (x,\xb)\; T_{-1/2} (y,\yb) } =} 
\nonumber \\
&=& \frac{-p_1^2}{2} \left( \frac{1}{w-x} + \frac{1}{w-y} \right) + \frac{p_1p_2}{w-v} 
\frac{(v-x)^{1/2}(v-y)^{1/2}}{(w-x)^{1/2}(w-y)^{1/2}} \; .
\eeq
Now, we define $p_{i\, l}$ to be the U(1) charge vector w.r.t.\ the left moving, i.e.\ holomorphic
U(1) currents, $p_{i\, r}$ the U(1) charge vector for the right moving, i.e.\ antiholomorphic U(1) currents.
Together with the analogous results for the derivatives with respect to  $v$, $\wb$ and $\vb$,
one gets by integration
\beqn
\lefteqn{\!\!\!\!\!\!\!\!\!\!\!\frac{\corr{V^t_{p_1} (w,\wb)\; V^t_{p_2} (v,\vb)\; T_{1/2} (x,\xb)\; 
T_{-1/2} (y,\yb) }}
{\corr{T_{1/2} (x,\xb)\; T_{-1/2} (y,\yb) }} =} \nonumber \\
&=& c((x-y),(\xb-\yb)) \; ((w-x)(w-y))^{-p_{1l}^2/2} ((v-x)(v-y))^{-p_{2l}^2/2} \nonumber \\
&& \times ((\wb-\xb)(\wb-\yb))^{-p_{1r}^2/2} ((\vb-\xb)(\vb-\yb))^{-p_{2r}^2/2} \nonumber \\
&& \times \left((x-y) \frac{(w-x)^{1/2}(v-y)^{1/2}+(w-y)^{1/2}(v-x)^{1/2}}
{(w-x)^{1/2}(v-y)^{1/2}-(w-y)^{1/2}(v-x)^{1/2}}  \right)^{-p_{1l}.p_{2l}}  \nonumber \\
&& \times \left((\xb-\yb) \frac{(\wb-\xb)^{1/2}(\vb-\yb)^{1/2}+(\wb-\yb)^{1/2}(\vb-\xb)^{1/2}}
{(\wb-\xb)^{1/2}(\vb-\yb)^{1/2}-(\wb-\yb)^{1/2}(\vb-\xb)^{1/2}}  \right)^{-p_{1r}.p_{2r}} 
\label{verkorr1} \\
&=& c((x-y),(\xb-\yb)) \; ((w-x)(w-y))^{-p_{1l}^2/2} ((v-x)(v-y))^{-p_{2l}^2/2}\nonumber \\
&& \times ((\wb-\xb)(\wb-\yb))^{-p_{1r}^2/2} ((\vb-\xb)(\vb-\yb))^{-p_{2r}^2/2} \nonumber \\
&& \times \left(\frac{((w-x)^{1/2}(v-y)^{1/2}+(w-y)^{1/2}(v-x)^{1/2})^2}
{(w-v)}  \right)^{-p_{1l}.p_{2l}}  \nonumber \\
&& \times \left(\frac{((\wb-\xb)^{1/2}(\vb-\yb)^{1/2}+(\wb-\yb)^{1/2}(\vb-\xb)^{1/2})^2}
{(\wb-\vb)}  \right)^{-p_{1r}.p_{2r}} \; . \label{verkorr2}
\eeqn

In this article, we are only interested in the two cases $p\equiv p_1= \pm p_2$.
For the case $p \equiv p_1=-p_2$ the correlator has to behave as the OPE
of two vertex operators in the limit  $w\rightarrow v$, i.e. as $(w-v)^{-p^2}$.
This fixes the coefficient to
\beq
c_-((x-y),(\xb-\yb)) = 4^{-p_r^2-p_l^2} \; .
\eeq
For the second relevant case $p \equiv p_1=p_2$, one fixes the coefficient by
continuation of the correlation function on the two sheeted Riemannian surface to
\beq
c_+((x-y),(\xb-\yb)) = 4^{-p_r^2-p_l^2} \; (x-y)^{2\,p_l^2} \; (\xb-\yb)^{2\,p_r^2} \; .
\eeq

One can check this result by noticing that the correlator \mref{verkorr1} shows the required
monodromy behaviour if one takes a vertex operator around a twistfield. This just
interchanges the sign of one charge $p_i$.

\end{appendix}


\bibliographystyle{JHEP}
\bibliography{holgerJHEP}

\end{document}